# A Framework for Simulating Real-world Stream Data of the Internet of Things

Weirong Xiu[1,2], Baozhu Li[3*], Xusheng Du[4*], Zheng Chu[4*]

[1] School of Information Technology and Engineering, Guangzhou College of Commerce, Guangzhou, China

[2] Management & Science University, Shah Alam, Selangor Darul Ehsan, Malaysia.

[3] Internet of Things & Smart City Innovation Platform, Zhuhai Fudan Innovation Institute, Zhuhai 519031, China

[4] School of Information Science and Engineering, Xinjiang University, Urumqi 830046, China

`xiuweirong@vip.163.com, tiger_1984@yeah.net,{chuzheng, duxusheng}@stu.xju.edu.cn`

**Abstract.** With the rapid growth in the number of devices of the Internet of Things (IoT), the volume and types of stream data are rapidly increasing in the real world. Unfortunately, the stream data has the characteristics of infinite and periodic volatility in the real world, which may leads to exceptions with the inefficient stream processing tasks. In this study, we report our recent efforts on this issue, with a focus on simulating stream data. Firstly, we explore the characteristics of the real-world stream data of the IoT, which helps us to understand the stream data in the real world. Secondly, the pipeline of simulating stream data is proposed, which can accurately and efficiently simulate the characteristics of the stream data to improve efficiency for specific tasks. Finally, we design and implement a novel framework that can simulate various stream data for related stream processing tasks. To verify the validity of the proposed framework, we apply this framework to stream processing task running in the stream processing system. The experimental results reveal that the related stream processing task is accelerated by at least 24 times using our proposed simulation framework with the premise of ensuring volatility and trends of stream data.

## 1 Introduction

In recent years, as the IoT [1] era has come, a variety of devices of the IoT continue to grow rapidly, including vehicle devices [2], mobile devices [3], smart furniture [4] and so on. These devices generate large and diverse types of stream data, such as social information, news feeds, sensor data, traffic information and application logs. On the one hand, the stream data are massive, diverse, rapidly-changing, and continuously generated in the real world. On the other hand, the business scenarios that require the processing of stream data continue to increase due to the rapid increase in the amount and type of stream data. It is an important task for analyzing or retrieving high-value information to simulate the stream data in the real world.

In order to efficiently and quickly process the stream data to extract useful information, developers need to develop programs to perform related tasks. During development and testing, the stream data is generated by two ways: (1) randomly generating stream data to simulate the real-world stream data [5-8], and (2) extracting the stream data in the real world in small or large quantities. Unfortunately, these ways have big problems. For example, randomly generating or extracting a small amount of real stream data cannot simulate all the characteristics of stream data, so the program cannot test all the exceptions and errors during code testing. And it will take a long time to perform this task by extracting a large amount of real-world stream data. In addition, performing tasks of load testing or parameter optimization will also encounter such problems. In particular, if the stream data

---

* Corresponding Baozhu Li
* Corresponding Xusheng Du
* Corresponding Zheng Chu
Zheng Chu, Xusheng Du and Baozhu Li contributed equally to this work.

This work is supported by the National Natural Science Foundation of China (Grant No. 61901191), the Shandong Provincial Natural Science Foundation (Grant No. ZR2020LZH005), China Postdoctoral Science Foundation (Grant No. 2022M713668), Educational Research Program of Guandong Province(Grant No. 2021KTSCX150),  and Scientific Research Foundation of Guangzhou(Grant No.202201011667).

in the real world is not used for load testing, the test results cannot fully reflect the characteristics of the business scenario, such as performance trend and the volatility of the stream data. In summary, the deadliest drawback is that the stream data generated by random manner or extracted part of the real-world stream data cannot fully reflect the trend and volatility of the stream data in the real world. This means that the efficiency of stream processing tasks associated with stream data of IoT is greatly reduced, further hindering the efficient development of IoT applications.

The purpose of this paper is to explore the characteristics of the stream data of the IoT in the real world and accurately simulate the characteristics of the stream data. Further, using the simulation framework proposed effectively improves efficiency for the stream processing task while preserving the trend and volatility of the stream data.

After studying the stream data of the IoT in the real world and summarizing the regularity of the corresponding stream data, a simulation pipeline for the stream data is proposed. The pipeline first performs preprocessing operations on the original stream data to identify the time information (i.e., timestamp or accurate time), and then normalize the stream data to shorten the scale of the data. At the same time, the stream data is sampled to reduce the amount of data. Finally, a production algorithm is proposed to send the sampled stream data. Especially, we design and implement a novel framework using the proposed simulation pipeline to simulate different kinds of stream data.

To the best of our knowledge, this work is the first that discuss the simulating the stream data of the IoT in the real world. Also, we propose a simple, accurate, and efficient pipeline that can better ensure the volatility and the trend of the original stream data and it can effectively reduce the time of data transmission to simulate stream data of the IoT in the real world.

In this paper, the main contributions of the current work include:

(1) We study the characters of real-world stream data in volatility and trend.

(2) We present a detailed pipeline, which contains three phrases including preprocessing, normalizing & sampling, and producing, to simulate stream data in the real world.

(3) We design and implement a framework, which can perform different simulation tasks for various types of stream data, using our proposed simulating pipeline.

(4) We perform comprehensive performance experiments to verify the validity of our framework, including analyzing volatility, impact on stream processing system, and effectiveness of our simulation framework. Through analyzing the experimental results, we find that our proposed simulation framework can achieve significant effects on three open data sets.

The remainder of this paper is structured as follows. In Section Ⅱ, we study the stream data of the IoT in the real world and then describe the simulation pipeline for the stream data in detail in Section Ⅲ. The framework for simulating the stream data is described in Section Ⅳ. We perform our experimental evaluation in Section Ⅴ. Lastly, Section Ⅵ makes a conclusion about our research.

## 2  Data In The Real World

Real-world stream data of the IoT is characterized by real-time, volatility and infinity. The stream data is a type of data that will not stop once it is generated. Once the data has flowed, it cannot be acquired again. Simultaneously, the stream data is sequential in the time dimension when being processed. There are two types of stream data: the unbounded stream data and the bounded stream data.

**Definition 1** (Unbounded stream data).

Unbounded stream data is defined as an infinite sequence, $U = s_1, s_2, \cdots, s_n, \cdots$. An element $s_i$ presents a tuple $<X_i, t_i>$ in the sequence, $s_i =<X_i, t_i>$. A timestamp $t_i$ is a positive integer value and $X_i$ is a data record with multidimensional properties in $s_i$. In addition, tuples arrive in chronological order, such as for any $i < j$, a tuple $s_i =<X_i, t_i>$ arrives earlier than $s_j =<X_j, t_j>$.

**Definition 2** (Bounded stream data).

Bounded stream data is defined as a finite sequence, $B = s_1, s_2, \cdots, s_n$. The definition of bounded stream data is the same as the unbounded stream data regarding tuples. Although the stream data in the real world is infinite, it can be composed of an infinite number of bounded stream data sequences, such as $U = B_1, B_2, \cdots, B_n, \cdots$. $B_n$ is bounded stream data of a fixed length of time, such as user behavior collected by mobile in an e-commerce website in one day or traffic information generated by vehicle device in one day.

It is an important mean for understanding the real-world stream data of the IoT to study real and open stream datasets.

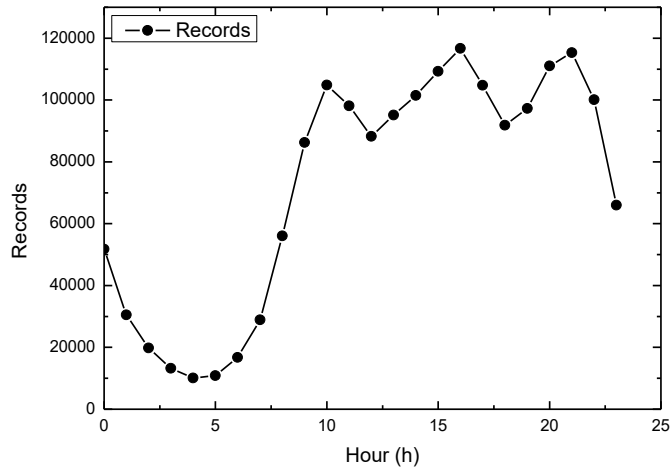

**Fig. 1.** The volatility and trend of SogouQ

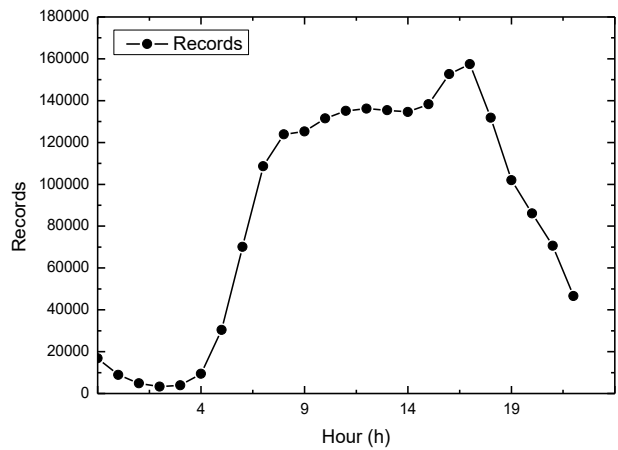

**Fig. 2.** The volatility and trend of Traffic

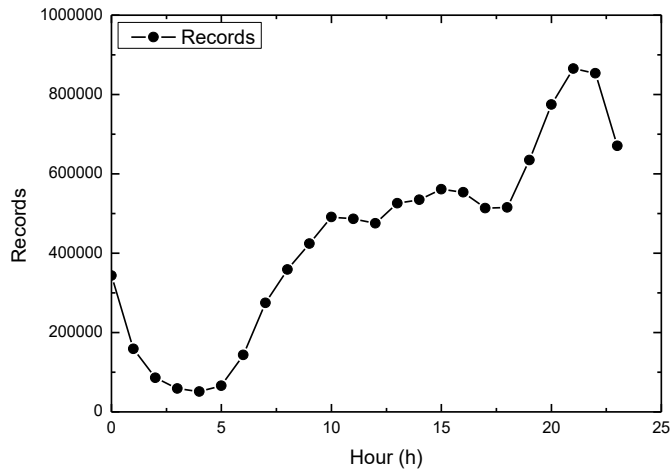

**Fig. 3.** The volatility and trend of UserBehavior

Fig. 1, Fig. 2 and Fig. 3 show the trend and volatility of the real-world stream data sets of the IoT, namely SogouQ [9], Traffic [10], and UserBehavior [11], by analyzing and drawing each data set separately. Due to the user's behavioral characteristics, the amount of data presents a large fluctuation in the day on three stream data sets. Therefore, stream data of the IoT in the real world is a sequence with the trend of large fluctuation and different kinds of stream data has different characteristics.

## 3 Using MS Word The Pipeline Of Simulating Stream Data

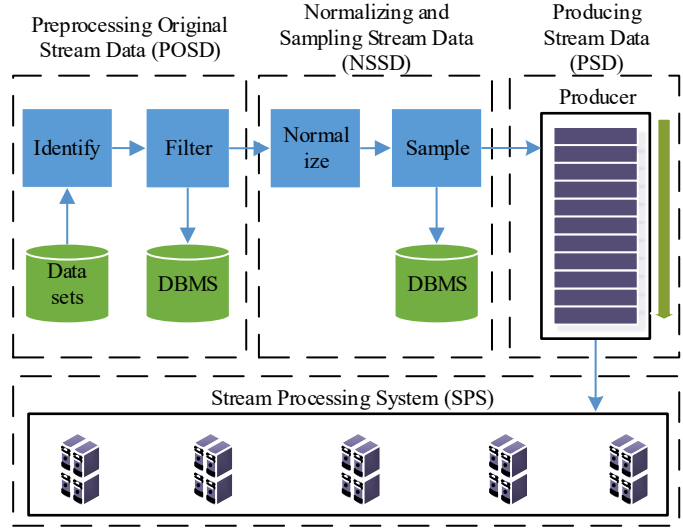

**Fig. 4.** Overview of simulation pipeline

Fig. 4 shows the pipeline of simulating real-world stream data of the IoT, which consists of four key components: Preprocessing original Stream Data (POSD), Normalizing and Sampling Stream Data (NSSD), Producing Stream Data (PSD), and Stream Processing System (SPS). The main job of POSD are to preprocess the original stream data set. The NSSD normalized the stream data after preprocessing and then performs sampling operations. The PSD sends the normalized and sampled stream data to the SPS to simulate the stream data of the IoT in the real world. Finally, the tasks in the SPS receive the real-world stream data simulated of the IoT and perform specific tasks.

### 3.1 Preprocessing

If the original stream data of the IoT is to be simulated into stream data in a smaller time range, each record in the original stream data needs to have information that can represent a specific time, such as a timestamp or an accurate time (i.e., YYYY-MM-DD HH:MM:SS). The POSD can convert it to a timestamp if it is a specific time. Therefore, identifying the timestamp or accurate time in the original stream data is the most important job in the preprocessing phase. In addition, some stream data use timestamps in different time zones such as UserBehavior [11], which requires timestamps using different time zones to be converted into the ones that using the same time zone. Considering that preprocessing for a specific stream data is an one-time job, so stream data preprocessed is stored in the database.

### 3.2 Normalizing And Sampling

Data normalization is an effective way to reduce the time range. Generally, there are many normalization methods such as Min-Max normalization [12], Z-score normalization [13], Sigmoid normalization [14], and so on. Z-score normalization normalizes the original data to a range of mean zero and variance one. Min-Max normalization linearly transforms the original data, which can maintain the relationship between the original records as shown formula 1:

$$t_{norm} = \frac{t - t_{min}}{t_{max} - t_{min}}(max - min) + min \tag{1}$$

where $min$ and $max$ represent the minimum and maximum timestamps of records in the original stream data, $t$ is the timestamp of the current tuple in the stream data, $max$ and $min$ represent the normalized maximum and minimum timestamps defined by user, and $t_{norm}$ (namely "scale_stamp") represents the normalized timestamp of the current record. We choose the Min-Max normalization to normalize the original stream data so that the data is dependent on the time series. Simulating the original stream data to a specified time range (i.e., 10 minutes, 20 minutes and so on) will lead to another problem, which is that the amount of the stream data is greater than the original data at each time point. That is to say, the original stream data is simulated into a small time range, but in fact, the amount of data at a certain point time will increase in an equivalent manner to the normalization multiple. Such results are unreasonable for simulating real-world stream data of the IoT. We will use the sampling method to reduce the amount of stream data normalized.

There are many common sampling methods, including random sampling, hierarchical sampling, cluster sampling, and systematic sampling. Random sampling has strong randomness and cannot reflect the underlying regularity of the stream data. Hierarchical sampling extracting a certain amount of samples from different layers can better make the sample representative, but the stream data needs to have a dependency on time series. The disadvantage of cluster sampling is that the error is too large. System sampling, finally, is selected to sample the stream data. The normalized stream data is sampled by setting a second as the distance so that the sequence of the stream data in time can be guaranteed. By using the above method, the amount of data per second can be reduced to the level of the original stream data amount. Stream data after normalizing and sampling will still be stored in the database because repeated normalizing and sampling operations are not performed. This will saves time and facilitates subsequent repeated testing. The normalizing and sampling processes are shown in Algorithm 1 (NSA).

---

**Algorithm 1:** NSA (Normalizing and Sampling Stream Data)

**Input**: $B$, $max$
$B = s_1, s_2, \cdots, s_n$: Original stream data to normalize and sample.
$max$: Time range defined by user.
**Output**: $D_s$
$D_s$: The simulated stream data.

---

Normalizing original stream data.
**For** $S_i$ in $B$ **do**
    $s_i.scale\_stamp \leftarrow (s_i - B_{min}) / (B_{max} - B_{min}) \, max$
    where $min$'s default value is 0, so
    $(max - min) + min = max$
**End For**
Sampling normalized stream data.
$multiple \leftarrow \text{Len}(B)/max$
**For** $i \leftarrow 0$ to $max$ **do**
    $block = B[scale\_stamp == i]$
    $rs = \text{Len}(block)/multiple$
    **For** $s_i$ in $block$ **do**
      **If** $i > rs$ **then**
        remove($s_i$)
      **End If**
    **End For**
**End For**
**Return** $D_s \leftarrow B$

---

The NSA mainly includes two processes: normalizing stream data and sampling stream data. In the first process, the NS calculated the scale_stamp defined by the user through the formula. 1 for every tuple Si. In the next process, the NS remove extra stream data by $multiple$ to reduce the amount of stream data.

## 3.3 Producing

Producer sends normalized and sampled stream data to the SPS in chronological order. The stream data cannot be transmitted in the order of iteration but should be able to simulate the characteristics of the stream data of the IoT in the real world when being send. Therefore, the program uses a parallel algorithm to send stream data in seconds to simulate the stream data, as shown in Algorithm 2 (PSDA).

---

**Algorithm 2:** PSDA (Producing Stream Data)

**Input**: $max$, $p$
$max$: Time range defined by the user.
$p$: Producer.
**Output**: $status$
$status$: The status of producing stream data (success:0, fault:1).

---

Loading stream data from DBMS.
$S \leftarrow \text{loaddatafromDBMS}(max)$;
Defining timer and bingding emit function.
$timer \leftarrow \text{threading.Timer}(emit, [0])$
Detecting lived emit thread.
**While** TRUE **do**
    Time.sleep(1);
    **If** threading.activeCount()<=0 **then**

>           Timer.cancel();
>           **Return** *status* ←0
>       **End If**
>   **End While**
>   Defining the thread of emitting stream data.
>   **Def** emit(*ite*)
>       *i*←*ite*
>       time.sleep(1)
>       Starting next thread to emit stream data.
>       *timer* ← threading.Timer(1.0, emit, [*ite* + 1])
>       Emitting stream data to stream processing system.
>       *block* ← *S*.loc ["scale_stamp"] == *i*
>       **if** *len(block) != 0* **then**
>           P(*block*)
>       **End If**
>   **End Def**

The PSDA first loads the stream data from the database based on the user-defined time range *max*, and then sends the stream data to the SPS in seconds via the producer in a parallel manner. Note that the variable "scale_stamp" is a normalization stamp (similar to a timestamp) generated during the normalization process to represent the chronological order after the normalizing.

## 4  The Framework For Simulating Stream Data

We present a novel framework for simulating stream data in the real world, which solves the problem we described earlier. Especially, the framework provides the general method to simulate various types of real-world stream data.

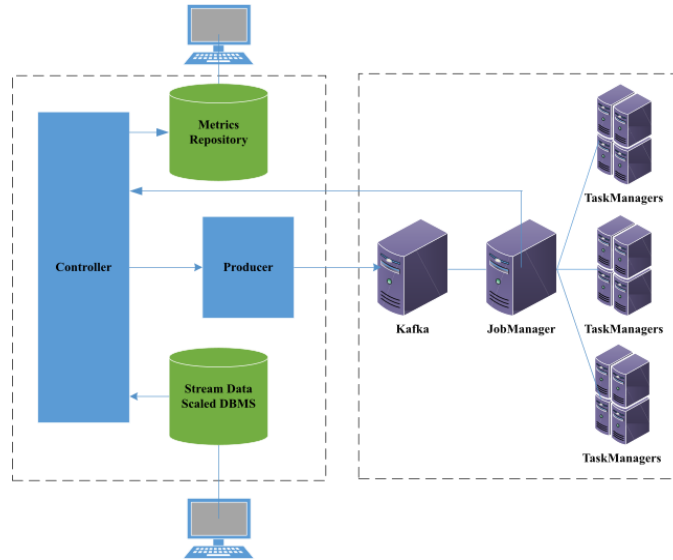

**Fig. 5.** Overview of framework

Fig. 5 shows an overview of the framework using the proposed pipeline, which comprises two parts. The first part is primarily the controller on the user side, which interacts with the stream processing system. It uses public API (such as REST API) to collect the metrics information of the stream processing system at runtime and save these metrics information to the metrics repository for viewing. These metrics include physical information, and performance of the streaming processing system. The second part is the stream processing system. In particular, the system includes a Kafka [15] node, which serves as an intermediate pipe that receives stream data sent from the producer at the user side and provides simulated stream data for the stream processing system. The stream processing system continuously reads stream data from Kafka to perform real-time processing tasks.

The controller is the core component in the user side, which controls the whole logics. The controller has the following three functions:

(1) Controlling the producer to load the stream data of user-defined time range from database for simulation;

(2) Collecting physical information of the workload and metrics information of the software system in the stream processing system;

(3) Managing metrics information of different stream data for viewing.

The stream processing system is a distributed computing cluster built by stream processing software (i.e. Apache Flink [16]).

In particular, the framework has three advantages:

(1) Independence: The framework does not depend on other platforms, it only needs a database to store original stream data and processed stream data;

(2) Universality: The framework can simulate a wide variety of stream data that must have timestamps or accurate time information;

(3) Traceability: Both original stream data and simulated stream data are stored in the database, which facilitates exception tracking and reuse.

## 5 Evaluation

### 5.1 Evaluation Methodology

The framework processes the three real-world stream data sets of the IoT (SogouQ, Traffic, and UserBehavior) using the pipeline described in Section 3, and then sends the stream data to the stream processing system (Apache Flink [16]) by the IoT devices. In the simulation process, the proposed simulation pipeline is evaluated by analyzing the volatility and trend of original stream data and the processed stream data. When normalizing, the original stream data of one day are simulated as stream data in one hour (i.e., 10 minutes to 60 minutes). This can greatly reduce the test time and better reflect the fluctuation characteristics existing in the original stream data.

**Data sets**: To mine the potential regularity of the stream data in the real world and evaluate the efficiency of the simulating pipeline on the data sets. We performed our proposed simulating pipeline on three stream data sets in the real world:

(1) SogouQ(SogouQ) [9]: The data set is a query log of the search engine, which includes a log of the Sogou search engine's query and user clicks in June 2008. The user behavior data contained in the data set provides the basic data for the researcher.

(2) Query Sub-dataset(Traffic) [10]: This sub-dataset was collected in Beijing, China between April 1, 2017, and May 31, 2017, from the Baidu Map. The query sub-dataset contains about 114 million user queries, each of which records the starting timestamp, coordinates of the starting location, coordinates of the destination, estimated travel time.

(3) User Behavior Data from Taobao for Recommendation(UserBehavior) [11]: User Behavior is a data set of user behaviors from Taobao, for recommendation problem with implicit feedback. The data set is offered by Alibaba.

### 5.2 Volatility Analysis

As mentioned in section 3, simulation for stream data is an important way to improve test efficiency and perform other tasks related to stream processing. To evaluate the quality of the simulating pipeline, it is necessary to compare the volatility of the original stream data with the simulated stream data. So, three metrics were selected: (1) *Average*, (2) *Variance*, and (3) *StandardVariance*, as shown in formula 2, 3, and 4.

$$Average = \frac{1}{max}\sum_{i=1}^{max} q_i \qquad (2)$$

$$Variance = \frac{1}{max}\sum_{i=1}^{max}(q_i - Average) \qquad (3)$$

$$StandardVariance = \sqrt{\frac{1}{max}\sum_{i=1}^{max}(q_i - Average)} \qquad (4)$$

where $max$ is the time range, $q_i$ is amount of stream data per second.

Table 1. Volatility stream data on SogoQ

| Time Range (s) | Average | Variance | Standard Variance |
|---|---|---|---|
| 600 | 25.35 | 234.15 | 15.30 |
| 1200 | 25.36 | 234.22 | 15.30 |
| 1800 | 25.37 | 234.42 | 15.31 |
| 2400 | 25.39 | 234.97 | 15.33 |
| 3000 | 25.38 | 235.16 | 15.34 |
| 3600 | 25.40 | 235.28 | 15.34 |

NOTE: ORIGINAL TIME RANGE OF STREAM DATA SET IS 86400S.

**Table 2.** Volatility stream data on Traffic.

| Time Range (s) | Average | Variance | Standard Variance |
|---|---|---|---|
| 600 | 21.41 | 113.02 | 10.63 |
| 1200 | 21.45 | 113.34 | 10.65 |
| 1800 | 21.46 | 113.36 | 10.65 |
| 2400 | 21.46 | 113.37 | 10.65 |
| 3000 | 21.45 | 113.61 | 10.66 |
| 3600 | 21.47 | 113.53 | 10.65 |

NOTE: ORIGINAL TIME RANGE OF STREAM DATA SET IS 86400S.

**Table 3.** Font sizes of headings Volatility stream data on UserBehavior.

| Time Range (s) | Average | Variance | Standard Variance |
|---|---|---|---|
| 600 | 121.90 | 4552.01 | 67.47 |
| 1200 | 122.01 | 4542.98 | 67.40 |
| 1800 | 122.04 | 4544.19 | 67.41 |
| 2400 | 122.06 | 4542.44 | 67.40 |
| 3000 | 122.07 | 4546.97 | 67.43 |
| 3600 | 122.08 | 4543.38 | 67.40 |

Note: original time range of stream data set is 86400s.

The results of Table 1, Table 2, and Table 3 show the stream data volatility, comparing the original stream data volatility characteristics with stream data after performing simulation at different time ranges on the three data sets.

**SogouQ**: The results of Table 1 show that the volatility metrics of the stream data at six time ranges (600s, 1200s, 1800s, 2400s, 3000s, 3600s) are close. By comparing the volatility of the time range of 3600s with the original stream data, it is found that: (1) the average value differs by about 1; (2) The variance value differs by approximately 4; (3) The standard variance value differs by approximately 1. In addition, with the time range becomes small, the volatility value becomes larger, comparing the simulated stream data with the original stream data volatility.

**Traffic**: as shown in Table 2, the volatility metrics of the stream data at the six time ranges are similar from the volatility metrics of original stream data. Reporting to the quality of the SogouQ, the simulation quality of the Traffic is better than the quality of the SogouQ.

**UserBehavior**: The results of Table 3 show that UserBehavior using different time ranges compare the volatility of the original stream data. On the volatility metrics, it can be found that the volatility after simulating can still show close to the volatility characteristics of the original stream data. Because the amount of stream data are large in every second, the UserBehavior are not better than SogouQ and Traffic regarding the three metrics.

By the comprehensive analysis of the simulation pipeline on three different stream data sets, it is clear that the volatility of simulated stream data is smaller than the volatility of the original stream data with the increase of the amount of data on the stream data sets.

### 5.3 Producing The Impact On Stream Processing System

Besides the statistic and analysis of the volatility of simulation pipeline on stream data in a static environment, it is important for evaluating the impact of simulation pipeline on stream processing system to perform stream processing task at runtime. The results of Fig. 6 show the impact of original stream data and simulated stream data on the stream processing system regarding network transmission.

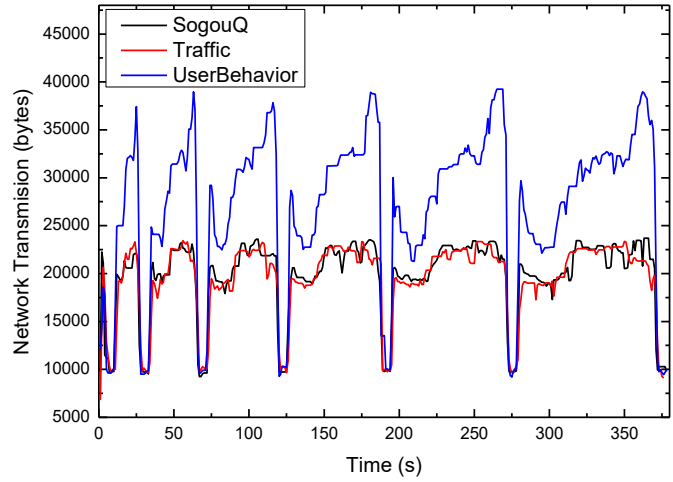

**Fig. 6.** The impact of network transmission on stream processing system

The results of Fig. 6 show the metrics of network transmission bytes with continuous six time ranges (10 min, 20 min, .., 60 min). They have shown similar volatility as a whole. The similar trend of them at six different time ranges can be seen from the figure. This is consistent with the results predicted by the simulation pipeline, indicating that the simulation pipeline can make the simulated stream data representative of the original stream data.

### 5.4 The Efficiency Comparison

The primary purpose of simulating stream data is to reduce test time, but the simulation process also takes time. It is, therefore, necessary to compare the efficiency of each data set in the simulation process.

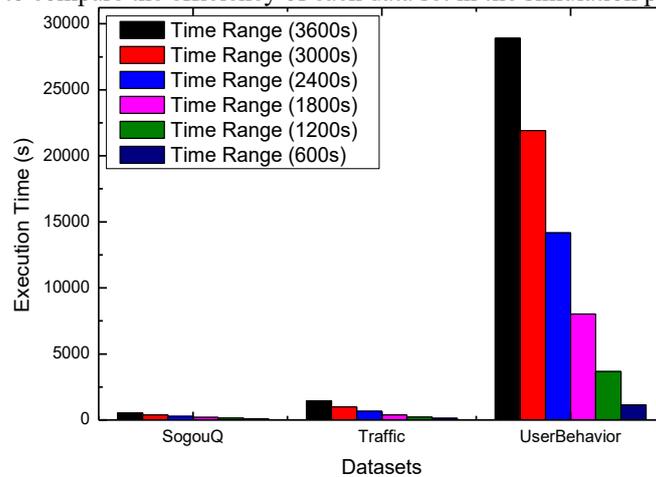

**Fig. 7.** The efficiency comparison of the different time ranges on three stream data sets

The results of Fig. 7 show the time spent by the simulation process on different data sets. With the time range of simulation continues to become small, the all three stream data sets show a decreasing trend regarding execution time. This phenomenon is because the amount of stream data is decreasing. The detail results obtained from this experiment was summarized in Table. 4.

**Table 4.** Font sizes of headings Volatility stream data on UserBehavior.

| Time Range (s) | SogouQ | Traffic | UserBehavior |
|---|---|---|---|
| 3600 | 536 | 1441 | 28906 |
| 3000 | 391 | 990 | 21918 |
| 2400 | 284 | 652 | 14161 |
| 1800 | 198 | 392 | 8029 |
| 1200 | 138 | 227 | 3676 |
| 600 | 94 | 126 | 1136 |

Note: Original time range of stream data set is 86400s.

With the time range of simulation increasing, the time spent on the simulation process is strongly dependent on the amount of data in the original stream data set. The amount of data of SogouQ, Traffic and UserBehavior increases in turn, and the time consumed by simulation also tends to increase.

By analyzing the impact of the time range and the amount of original data sets on execution time, the time range of simulation should be small as much as possible with the premise of ensuring the volatility of stream data, improving efficiency in the simulation process.

## 6  Conclusion

In this dissertation, we study the volatility and the trend of the stream data of IoT in the real world. The stream data in the real world have the potential regularity of periodicity after researching. To solve the inefficiency problem of a large amount of stream data in code debugging, load testing and parameter optimization, a simulation framework using our proposed pipeline to address these problems is proposed. Through experiments and analysis, the pipeline can well maintain the characteristics of the volatility and trends of the original stream data. At the same time, we compare the impact of different time ranges on the stream processing system and verify that the simulation can effectively improve the efficiency of stream processing in the experiment. Especially, if the time range is set to less one hour (3600s), the related stream processing tasks are accelerated by at least 24 times with the premise of ensuring volatility and trends of stream data.

## 7  Acknowledgement

This work was supported in part by the National Natural Science Foundation of China under Grant Nos. 61862060, 61462079, 61562086315 and 61562078, and the Doctoral Science and Technology Innovation Project in Xinjiang University under Grand No. XJUBSCX-201901.